\documentclass[journal]{IEEEtran}
\usepackage{cite}

\ifCLASSINFOpdf
\usepackage[pdftex]{graphicx}
\else
\fi
\usepackage{amsmath}
\usepackage{url}

\begin{document}
%
\title{Co-Designed Reflective and Leaky-Waveguide Low-Pass Filter for Superconducting Circuits}
%
%
%

\author{Linus~Andersson,
        Benjamin~Olsson,
        Simone~Gasparinetti
        and~Robert~Rehammar
\thanks{This work was supported by the KAW
via the Wallenberg LaunchPad (WALP) program. The work of S. Gasparinetti and R. Rehammar was supported by the Knut and Alice Wallenberg foundation via the Wallenberg Centre for Quantum Technology (WACQT). S. Gasparinetti acknowledges funding from the European Research
Council via Grant No. 101041744 ESQuAT. L. Andersson acknowledges
funding from the European Union via Grant
No. 101135240 JOGATE. \textit{(Corresponding author: Linus Andersson.)}}
\thanks{The authors was with the Department of Microtechnology and Nanoscience, 
Chalmers University of Technology, 412 96 Göteborg, Sweden (e-mail: anlinus@chalmers.se)} 
}

%
%

\markboth{}%
{Andersson \MakeLowercase{\textit{et al.}}: Co-Designed Reflective and Leaky-Waveguide Low-Pass Filter for Superconducting Circuits}
%



\maketitle

\begin{abstract}
A stepped-impedance low-pass filter with integrated hollow waveguide absorbers is presented. The filter combines low insertion loss in the passband with strong attenuation at high frequencies, making it well suited for superconducting quantum computing applications, where qubits are sensitive to both near-band and far out-of-band radiation. The structure is implemented in a rectangular coaxial geometry, with inductive sections coupled to circular hollow waveguides oriented orthogonally to the transmission axis. Above their cutoff frequency, these waveguides efficiently couple to radiation inside the stepped-impedance filter, absorbing energy that would otherwise cause Cooper pair breaking in conventional superconductors. Optimal dimensions were obtained using a differential evolution algorithm applied to interpolated electromagnetic simulation data. A prototype was fabricated and characterized using a calibrated vector network analyzer up to 67 GHz. Measurements confirm a 3 dB cutoff frequency at 13.5 GHz, insertion loss below 0.45 dB for frequencies under 8 GHz, and more than 52.7 dB rejection above 17.3 GHz. The design offers a compact, low-loss solution for near-band filtering and suppression of quasiparticle-generating radiation in cryogenic quantum systems.
\end{abstract}


%
\IEEEpeerreviewmaketitle

\section{Introduction}
%
%
%
%

\label{sec:introduction}
Superconducting circuits are among the most promising building blocks for implementing fully functional quantum computers \cite{devoret_superconducting_2013, bardin_quantum_2020}. Based on Josephson junctions \cite{VanDuzer1999}, these devices, in combination with passive microwave circuitry, form highly nonlinear quantum circuits that can be operated as qubits. Superconducting qubits are highly sensitive to electromagnetic radiation both near and far from their operational frequency, typically in the 4–8 GHz range. To mitigate the detrimental effects of such radiation, qubit input and output lines are commonly filtered using combinations of LC filters, absorptive filters and cryogenic isolators.

Near-band, off-resonant electromagnetic interference has been shown to adversely affect the performance of superconducing qubits. Known mechanisms include AC Stark shifts of qubit frequencies~\cite{schuster_ac_2005, wallraff_approaching_2005} and resonant excitation of higher-order circuit modes, box modes of the circuit housing package~\cite{huang_microwave_2021}, and spurious two-level systems and fluctuators in the vicinity of the qubits.
To mitigate these effects, filters with sharp rolloff at the edge of the operational band are typically used.

Another source of decoherence comes into play at much higher frequencies. In a superconductor well below its critical temperature, $T_C$, the dominant charge carriers are Cooper pairs \cite{M.Tinkham_intro_sc}. For frequencies above \(2\Delta/\hbar\), where \(\Delta\) is the superconducting energy gap and \(\hbar\) is the reduced Planck constant, radiation incident on the superconductor, typically above 80 GHz for aluminum thin films, can break the Cooper pairs, creating Bogoliubov quasiparticles \cite{M.Tinkham_intro_sc}. These quasiparticles can degrade the qubit lifetime and coherence by tunneling through the Josephson junction \cite{glazman_bogoliubov_2021, catelani_relaxation_2011, aumentado_quasiparticle_2023}. Despite quasiparticle densities being experimentally observed to be several orders of magnitude larger than predicted \cite{wang_measurement_2014}, the presence of quasiparticles is not expected to limit qubit performance until coherence times approach the millisecond range \cite{riste_millisecond_2013}. As coherence times now start to reach such levels \cite{somoroff_millisecond_2023, bland_2d_2025, tuokkola_methods_2025}, careful strategies to mitigate quasiparticle effects have become increasingly important.

The reasons for the excess of quasiparticles are thermal radiation inside the cryogenic environment \cite{gordon_environmental_2022, connolly_coexistence_2024}, ionizing radiation, generating bursts of quasiparticle tunneling events \cite{vepsalainen_impact_2020, mcewen_resolving_2022, cardani_reducing_2021, harrington_synchronous_2024, wilen_correlated_2021} and phonon-only events related to stress in the substrate and thin-films \cite{yelton_correlated_2025}. Mitigating sensitivity to ionizing impacts and phonon-only events are important, given their detrimental effects on quantum error correction \cite{mohseni_how_2025}. Resilience to such events can be addressed through gap engineering of the junctions \cite{mcewen_resisting_2024, martinis_saving_2021, aumentado_nonequilibrium_2004, nho_recovery_2025} and phonon-trapping techniques \cite{yelton_correlated_2025, riwar_normal-metal_2016, patel_phonon-mediated_2017}.

Reducing quasiparticle tunneling caused by high-frequency radiation from the cryogenic environment can be achieved through light-tight packaging and shielding of the quantum processor \cite{barends_minimizing_2011}, as well as by using infrared (IR) blocking filters on the input and output lines for a qubit. Usually, filters for this purpose are based on coaxial waveguides filled with absorptive materials \cite{halpern_far_1986, santavicca_impedance-matched_2008, paquette_absorptive_2022, martinis_experimental_1987, slichter_millikelvin_2009}. Measurements have shown to significantly reduce the average quasiparticle tunneling rates, as demonstrated in \cite{connolly_coexistence_2024} through improved light-tightness and extensive use of IR blocking filters.

While effective at absorbing high-frequency radiation, absorptive materials inside the waveguide also introduce unwanted loss at frequencies where low insertion loss is critical. On readout lines, where quantum-noise-limited parametric amplifiers \cite{roy_introduction_2016, aumentado_superconducting_2020} and cryogenic low-noise amplifiers \cite{pospieszalski_extremely_2005, zeng_sub-mw_2024} are used, preserving signal level is crucial. Therefore, there is a clear need for filter technologies that overcomes the trade-off of having low in-band loss while providing strong out-of-band absorption. On input lines that route control signals to the quantum processor, heavy attenuation is applied to thermalize room-temperature noise down to millikelvin temperatures \cite{krinner_engineering_2019}. Despite being heavily attenuated, maintaining a flat attenuation profile is also desired in order to avoid pulse distortions that can degrade qubit gate fidelity \cite{simbierowicz_microwave_2022}. Traditional absorptive filters often fail in this regard, owing to their strong frequency-dependent loss. 

Recently, a new absorptive low-pass filter type, termed High-Energy Radiation Drain (HERD), was demonstrated. This filter overcome this trade-off, exhibiting less than 0.15 dB insertion loss below 12 GHz while providing more than 60 dB absorption above 70 GHz \cite{rehammar_low-pass_2023}. Based on a coaxial waveguide, the HERD filter removes high-frequency radiation through strong coupling to hollow waveguides (HWs) implemented in the outer conductor. By setting the hollow waveguide cutoff frequency sufficiently high to avoid parasitic evanescent-mode tunneling for in-band signals, losses are mostly limited by conductive and dielectric losses in the filter and connectors and impedance matching. Since its demonstration, the HERD filter has been deployed in cryogenic systems achieving state-of-the-art qubit readout \cite{kurilovich_high-frequency_2025, hazra_benchmarking_2025}, high-saturation traveling-wave parametric amplification \cite{gaydamachenko_rf-squid-based_2025}, and millisecond qubit coherence times \cite{bland_2d_2025}. Quasiparticle tunneling rates comparable to the best achieved with conventional absorptive filters have also been demonstrated \cite{nho_recovery_2025}. However, a weakness of the conventional HERD design is its slow rolloff from passband to the highly attenuating stopband. Therefore, this filter is often combined with separate LC filter with lower cutoff frequency and sharper rolloff. 

In this paper, we report on investigations that extend this filtering approach by integrating the hollow waveguide structures into a stepped-impedance low-pass filter, achieving both high near-band rejection and strong absorption of Cooper pair-breaking frequencies well beyond the design reported in \cite{rehammar_low-pass_2023}. By optimizing the filter geometry, we achieve a 3 dB cutoff frequency at 13.5 GHz, an insertion loss of less than 0.45 dB below 8 GHz, and more than 52.7 dB of rejection above 17.3 GHz. This filtering technique eliminates the need for using combinations of LC and absorptive filters, reducing component overhead and enabling further scaling of quantum processors.

\section{Design}

\begin{figure}
  \centering
  \includegraphics[width=3.5in]{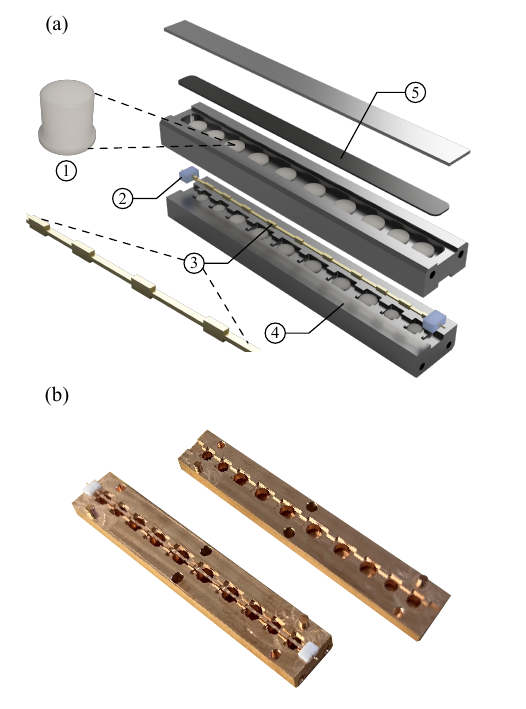}
  \caption{(a) A 3D exploded view of the filter assembly (colored for visibility). The different parts of the assembly are indicated with numbers. 1: The MACOR ceramic slab which are inserted into the HWs. 2: PTFE piece used to fixate the center conductor. 3: The center conductor where the modulation of the conductor height can be seen. 4: Outer conductor block with different sections of high and low impedance. 5: Absorptive foam to absorbs the leaking radiation from the HWs. (b) A photograph of the outer conductor blocks with the central line fixated to one of the halves with pieces of PTFE.}
  \label{filter_3d}
\end{figure}
\begin{figure}
  \centering
  \includegraphics[width=3.5in]{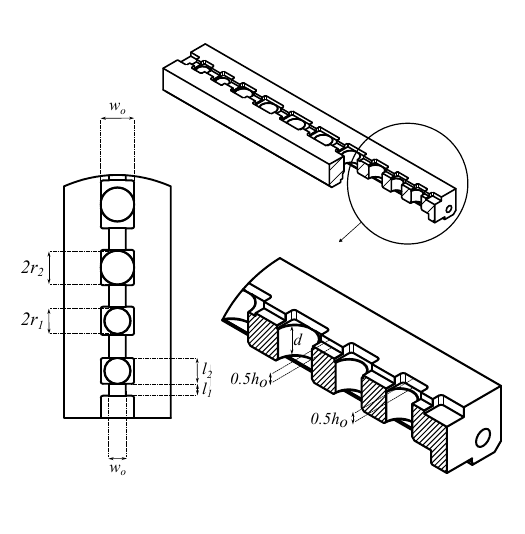}
  \caption{Cross-sectional view of the outer conductor block with dimensions.}
  \label{h2r_detailed_view}
\end{figure}
The stepped-impedance HERD low-pass filter is constructed from two Computer Numerical Control (CNC)-machined copper blocks that, when soldered together, form the outer conductor of a rectangular coaxial transmission line. A flat copper strip serves as the center conductor. This configuration supports a transverse electromagnetic (TEM) mode, as well as higher order transverse electric (TE) and and transverse magnetic (TM) modes. The center transmission line is coupled to circular HWs oriented perpendicularly out from the main waveguide's outer conductor, see~\cite{rehammar_low-pass_2023} for a detailed discussion. The HWs have a cutoff frequency of approximately 25~GHz, well above the typical operating frequency of superconducting qubits (4-8~GHz). An exploded view of the filter assembly can be found in Fig.~\ref{filter_3d} together with photographs of the manufactured prototype.

The filter comprises several sections of varying width and height for the inner and outer conductor. These variations create sections of high or low impedance. In general, if a section of transmission line is short compared to the wavelength, it can be approximated as a reactive element. For a high-impedance section, the corresponding inductance is approximately
\begin{equation}
    L = \frac{\beta \ell Z_h}{R_0},
\end{equation}
while for a low-impedance section, the equivalent capacitance is
\begin{equation}
    C = \frac{\beta \ell R_0}{Z_\ell},
\end{equation}
where $\beta$ is the phase constant, $\ell$ is the physical length of the section~\cite{pozar2012}. The impedances $Z_h$ and $Z_\ell$ are the characteristic impedance of the high- and low-impedance sections, respectively. $R_0$ is the system reference impedance. 

In our implementation, the impedance is modulated by adjusting the outer conductor width $w_o$ and height $h_o$, as well as the center conductor height $h_i$. The HWs are embedded in the high-impedance sections for two key reasons. Firstly, their presence locally increases the inductance of the section due to partial field leakage and evanescent mode excitation near the waveguide apertures. Second, a larger outer cross-section in the high-impedance regions permits full aperture exposure of the HWs, enhancing the electromagnetic coupling to out-of-band radiation. The HWs are symmetrically placed above and below the center conductor, coinciding with regions of maximum transverse electric field. A detailed view of the filter geometry can be seen in Fig. \ref{h2r_detailed_view}.

\subsection{Dimensions and Material Choices}

Achieving a high impedance contrast, $Z_h / Z_\ell$, is critical to produce a sharp transition between passband and stopband. In the present prototype, the low-impedance sections have $Z_\ell = 22.1~\Omega$, while the high-impedance sections have $Z_h = 85.9~\Omega$.

The TEM mode profiles are simulated in COMSOL Multiphysics~\cite{comsol}, where the characteristic impedance is calculated as
\begin{equation}
    Z_0 = \frac{V}{I} = \frac{\int_\ell \vec{E}_t\cdot d\vec{\ell}}{\oint_C \vec{H}_t\cdot d\vec{\ell}},
\end{equation}
where the numerator represents the line integral of the transverse electric field between the conductors and the denominator the loop integral of the transverse magnetic field around the center conductor. The impedance varies along the HW-loaded inductive sections; we extract $Z_0$ at the axial midpoint, where the HW aperture is fully exposed.

To reduce the cutoff frequency and thereby the physical size of the HWs, they are filled with the machinable glass ceramic MACOR. Two distinct slab radii, $r_1$ and $r_2$, are used, which further suppresses out-of-band re-transmission of the stepped impedance filter by introducing poles at different frequencies. Sufficient depth $d$ of the HWs are critical to prevent evanescent tunneling of the in-band modes. Below cutoff, the fields decay exponentially according to the expression
\begin{equation}
    F(d) = e^{-\gamma d}, \quad \gamma = \sqrt{k_c^2 - k_0^2},
\end{equation}
where $k_c$ is the cutoff wavenumber and $k_0$ is the vacuum wavenumber. Away from the center conductor, the HWs are terminated with a carbon-loaded polyethylene foam to absorb radiated power. The foam is held in place with a piece of metal. 

The center conductor is cut from a 0.5 mm copper sheet and slotted into a PTFE-filled 50~$\Omega$ TEM transmission section for alignment and mechanical stability. A summary of the geometrical parameters and material properties can be found in Table~\ref{tab:dimensions}. 
\begin{table}[h]
\centering
\caption{Filter Geometry Parameters}
\begin{tabular}{|l|c|c|}
\hline
\textbf{Parameter} & \textbf{Symbol} & \textbf{Value} \\ \hline
Inner conductor height & $h_i$ & 0.8, 0.3 mm \\ \hline
Outer conductor width & $w_o$ & 1.5, 3 mm \\ \hline
Outer conductor height & $h_o$ & 0.5, 1.45 mm \\ \hline
HW radii & $r_s$, $r_l$ &  1.15, 1.5 mm \\ \hline
HW depth & $d$ & 2.5 mm \\ \hline
MACOR permittivity & $\varepsilon_r$ & 5.64 \\ \hline
MACOR loss tangent & $\tan \delta$ & $2.5 \times 10^{-3}$ \\ \hline
\end{tabular}
\label{tab:dimensions}
\end{table}

\subsection{Optimization Procedure}
\label{sec:optimization}

In order to reach the desired response of the LC filter, the lengths of each low and high impedance section needs to be determined. For conventional, resonant filters, the optimal parameters can be synthesized from known polynomials~\cite{pozar2012}. As a result, only a final, local optimization of the physical geometry is usually needed to reach the desired response. However, for the filter prototype presented in this paper, this approach was found to be inadequate. The filter comprises 21 sections, yielding a $21^{\mathrm{st}}$-order stepped-impedance low-pass filter. Ten of those are the inductive sections, which embed the HWs. To account for the spatial variation of impedance within these sections caused by the presence of the HWs, as well as the complex field interactions at frequencies above cutoff, the inductive sections are modeled using COMSOL. This allows for accurate treatment of the geometry and electromagnetic response.
\begin{figure}
  \centering
  \includegraphics[width=3.5in]{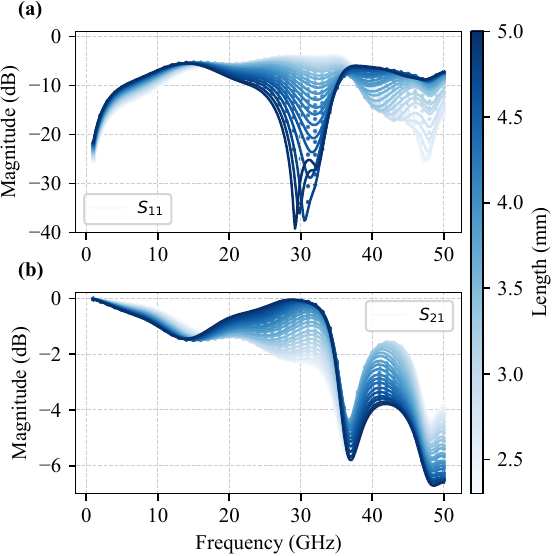}
  \caption{Simulated (a) reflection coefficient $S_{11}$ and (b) transmission coefficient $S_{21}$ as functions of frequency for varying inductive section lengths. The length is swept from 2.5 mm to 5.0 mm, as indicated by the color scale.}
  \label{fig:inductive_sparams}
\end{figure}

To optimize the length of the sections, we proceed as follows. First, we simulate the S-parameters of each inductive section as a function of physical length using frequency-domain finite element modeling (FEM) in COMSOL Multiphysics~\cite{comsol}. These results are then interpolated to obtain continuous S-parameter functions $S_{21}(f, \ell)$ and $S_{11}(f, \ell)$, seen in Fig.~\ref{fig:inductive_sparams}. From these results, their corresponding ABCD matrices are computed. The ABCD matrix for the capacitive sections are modeled using standard transmission line theory
\begin{equation}
    \mathbf{T}_{i} =
    \begin{bmatrix}
    \cos \beta \ell & j Z \sin \beta \ell \\
    j \frac{1}{Z} \sin \beta \ell & \cos \beta \ell
    \end{bmatrix},
\end{equation}
where the impedance comes from the mode analysis described above. To reduce the number of parameters in the optimization, the filter is made symmetric in the LC-distribution; therefore, only 11 parameters need to be optimized. However, the central (11th) parameter corresponds to a capacitive section, which is split into two separate matrices in the calculation. The total filter response is obtained by cascading ABCD matrices according to
\begin{equation}
    \mathbf{T_{\text{tot}}}(\omega; \mathbf{\ell}) = \left( \prod_{i=1}^{N/2} \mathrm{T_i(\omega, \ell_i)} \right) \left( \prod_{i=N/2}^{1} \mathrm{T_i(\omega, \ell_i)} \right)
\end{equation}
where N=22 and $T_i$ is inductive when $i$ is odd and capacitive when $i$ is even. 

To determine the optimal lengths, we use a differential evolution (DE) optimizer from the Python package SciPy \cite{2020SciPy-NMeth}, which is robust to local minima and does not require gradient information. The cost function penalizes passband ripple, insertion loss, and insufficient stopband attenuation. In this study, the optimization aimed for a return loss better than 20 dB below 12 GHz and an insertion loss greater than 50 dB in the 15–40 GHz range. As discussed in Section~\ref{sec:introduction}, superconducting qubits typically operate in the 4–8 GHz band. The 12 GHz cutoff was selected to also accommodate pump tones in the 10–12 GHz range, commonly used in driving parametric amplifiers for qubit readout. Additionally, we anticipate that the HW absorbers will dominate the stopband response above 40 GHz, which motivated choosing an upper frequency limit of 40 GHz in the optimization. Once convergence is reached, the full 3D model of the filter is simulated using FEM in COMSOL with all the optimized parameters to validate agreement with the simplified matrix-based model.

\section{Results}

In this section, the results from investigations of the full filter prototype are presented. The optimized parameters for the different sections can be found in Table~\ref{tab:opt_lengths}.

\subsection{Optimized Filter}

\begin{figure}
  \begin{center}
  \includegraphics[width=3.5in]{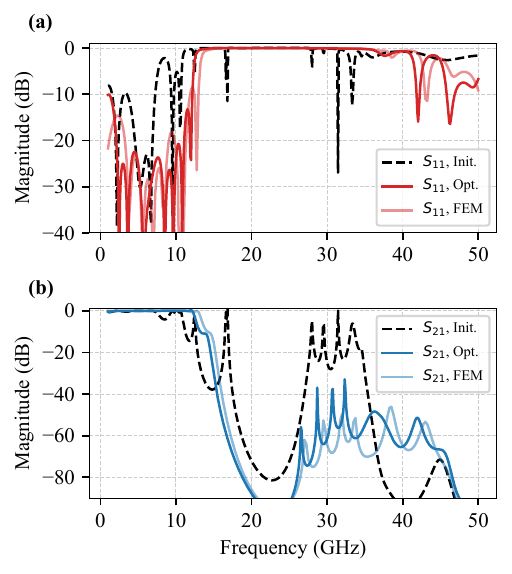}\\
  \caption{Simulated (a) reflection coefficient, $S_{11}$ and (b) transmission coefficient, $S_{21}$, showing the results of the filter geometry optimization. The dashed curves correspond to a randomly chosen initial parameter set. The solid curves represent the optimized solution, simulated either with the ABCD matrix model (darker shades of color) or with a full 3D simulation (lighter shades). The observed shifts in response are attributed to 50~$\Omega$ PTFE sections, which are included in the FEM model but not in the ABCD-matrix optimization. The close agreement between the optimization results and full-wave simulations highlights the accuracy of the proposed design method.
}\label{fig:sparam_optimization}
  \end{center}
\end{figure}

The optimization of the structure was performed using the method described in Section~\ref{sec:optimization}. Starting from random initial parameters, the optimizer converges to an acceptable solution, as shown by the simulated scattering parameters in Fig.~\ref{fig:sparam_optimization}. The return loss remains below -18 dB between 2.85-10.95 GHz, and the rejection in the stopband exceeds 50 dB above 15.85 GHz, close to the goals set in Section~\ref{sec:optimization}. The results from the optimization are then verified using a full 3D model of the complete filter. As seen, the ABCD-matrix model shows good agreement with the full 3D EM simulation.

\begin{table}[!t]
\centering
\caption{Optimized Section Lengths of the Filter Prototype}
\label{tab:opt_lengths}
\begin{tabular}{c c}
\hline
\textbf{Section} & \textbf{Length (mm)} \\
\hline
1 & 1.06 \\
2 & 2.30 \\
3 & 2.05 \\
4 & 2.54 \\
5 & 1.91 \\
6 & 3.16 \\
7 & 1.91 \\
8 & 4.28 \\
9 & 1.77 \\
10 & 4.63 \\
11 & 0.80 \\
\hline
\end{tabular}
\end{table}

\subsection{Measurements}

\begin{figure}
 \begin{center}
 \includegraphics[width=3.5in]{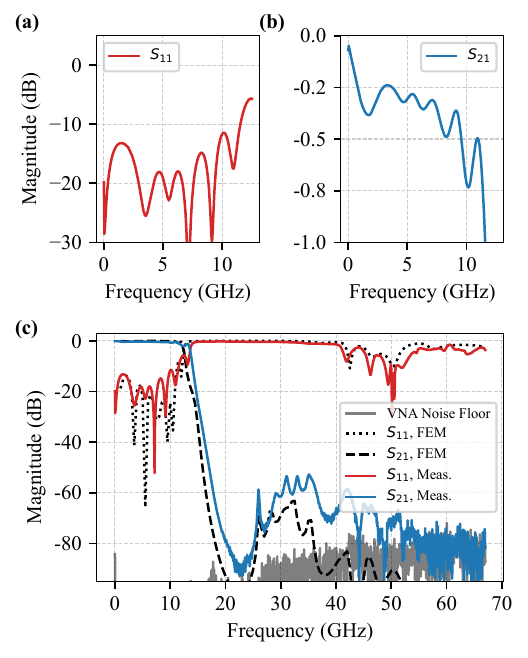}\\
 \caption{Simulated and measured results for the final filter design. (a) and (b) show the measured reflection coefficient $S_{11}$ and transmission coefficient $S_{21}$ in the passband, respectively. (c) shows the full-band response, comparing the measured data (solid red and blue) with full-wave FEM simulations (dashed black). The measured transmission closely follows the simulated response across the full frequency range. The light gray trace indicates the measured VNA noise floor using a short circuit termination. Minor deviations are attributed to connector transitions and modeling approximations outside the passband.}
  \label{fig:vna_measurement}
 \end{center}
\end{figure}

The filter prototype was characterized using a calibrated vector network analyzer (VNA) with a measurement range extending up to 67 GHz. To interface with the SMA connectors on the filter, 1.85 mm to 2.92 mm coaxial adapters were used in conjunction with 1.85 mm cables. In order to further suppress the unwanted re-transmission of the stepped-impedance filter, the apertures of the HWs on section 6, 8 and 10 was increased from 2.3 mm to 3 mm in diameter. The measured S-parameters are presented in Fig.~\ref{fig:vna_measurement}, along with simulated data obtained from a full-wave 3D electromagnetic model of the structure. Excellent agreement is observed between measurement and simulation across the whole passband, around the cutoff frequency, and at the onset of the stopband. Deviations above 30 GHz are attributed to limitations in the simulation model, in particular, the use of idealized scattering boundary conditions to terminate the hollow waveguide sections.

The VNA noise floor was determined by terminating one port with a short while operating at a 100 Hz IF bandwidth. As shown in the plot, the measured stopband approaches the noise floor around 55 GHz, indicating continued attenuation at higher frequencies. The final prototype exhibits a 3 dB cutoff frequency at 13.5 GHz, an insertion loss below 0.45 dB for frequencies under 8 GHz, and a rejection exceeding 52.7 dB above 17.3 GHz, demonstrating the effectiveness of the optimized filter geometry.

\section{Conclusion}


In this work, a compact stepped-impedance low-pass filter integrating hollow waveguide absorbers has been presented for suppressing high-frequency radiation in superconducting quantum processors. The design achieves a sharp passband-to-stopband transition with strong coupling to hollow waveguide terminations, which effectively absorb out-of-band radiation above the Cooper pair breaking frequencies of conventional superconducting thin films. By optimizing the filter dimensions using an ABCD-matrix-based approach guided by electromagnetic simulations, the prototype achieves a 3 dB cutoff at 13.5 GHz, an insertion loss below 0.45 dB for frequencies below 8 GHz, and rejection exceeding 52.7 dB above 17.3 GHz.

The filtering approach demonstrated here is general, and the underlying design principles can be adapted to different cutoff frequencies, depending on the available materials and fabrication processes. To design for other cutoff frequencies, changing the length of each section as well as the diameter of the hollow waveguides is required. We consider this filter to exhibit a Chebyshev-type response. Other filter responses may be possible with a different topology than stepped impedance.
Further enhancement of stopband rejection may be possible by including the hollow waveguide diameter as an additional optimization parameter, potentially introducing more poles near the re-transmission region of the stepped-impedance filter. This was not investigated in the present work due to added complexity in both optimization and fabrication.

These results highlight a compact filtering technique that combines multiple desirable properties well suited for the development of large-scale quantum computing architectures based on superconducting qubits. The proposed concept is also broadly applicable to superconducting circuits and other cryogenic or quantum systems where low insertion loss and a wide stopband are critical.

The filter is patent pending.


%



\section*{Acknowledgment}

The authors would like to thank Lars J\"onsson for manufacturing the prototype. L.A., R.R.~and S.G.~are co-founders and shareholders of Sweden Quantum AB. Other authors declare no competing interests.

\ifCLASSOPTIONcaptionsoff
  \newpage
\fi



%

\bibliographystyle{IEEEtran}
\bibliography{IEEEabrv, bibtex/bib/references_ieee_style}

\begin{thebibliography}{10}
\providecommand{\url}[1]{#1}
\csname url@samestyle\endcsname
\providecommand{\newblock}{\relax}
\providecommand{\bibinfo}[2]{#2}
\providecommand{\BIBentrySTDinterwordspacing}{\spaceskip=0pt\relax}
\providecommand{\BIBentryALTinterwordstretchfactor}{4}
\providecommand{\BIBentryALTinterwordspacing}{\spaceskip=\fontdimen2\font plus
\BIBentryALTinterwordstretchfactor\fontdimen3\font minus \fontdimen4\font\relax}
\providecommand{\BIBforeignlanguage}[2]{{%
\expandafter\ifx\csname l@#1\endcsname\relax
\typeout{** WARNING: IEEEtran.bst: No hyphenation pattern has been}%
\typeout{** loaded for the language `#1'. Using the pattern for}%
\typeout{** the default language instead.}%
\else
\language=\csname l@#1\endcsname
\fi
#2}}
\providecommand{\BIBdecl}{\relax}
\BIBdecl

\bibitem{devoret_superconducting_2013}
\BIBentryALTinterwordspacing
M.~H. Devoret and R.~J. Schoelkopf, ``Superconducting circuits for quantum information: An outlook,'' \emph{Science}, vol. 339, no. 6124, pp. 1169--1174, Mar. 2013. [Online]. Available: \url{https://www.science.org/doi/10.1126/science.1231930}
\BIBentrySTDinterwordspacing

\bibitem{bardin_quantum_2020}
\BIBentryALTinterwordspacing
J.~C. Bardin, D.~Sank, O.~Naaman, and E.~Jeffrey, ``Quantum computing: An introduction for microwave engineers,'' \emph{IEEE Microw. Mag.}, vol.~21, no.~8, pp. 24--44, Aug. 2020. [Online]. Available: \url{https://ieeexplore.ieee.org/document/9134701}
\BIBentrySTDinterwordspacing

\bibitem{VanDuzer1999}
T.~V. Duzer and C.~W. Turner, \emph{Principles of superconductive devices and circuits}, 2nd~ed.\hskip 1em plus 0.5em minus 0.4em\relax Upper Saddle River, NJ, USA: Prentice Hall, 1999.

\bibitem{schuster_ac_2005}
\BIBentryALTinterwordspacing
D.~I. Schuster \emph{et~al.}, ``Ac stark shift and dephasing of a superconducting qubit strongly coupled to a cavity field,'' \emph{Phys. Rev. Lett.}, vol.~94, no.~12, p. 123602, Mar. 2005. [Online]. Available: \url{https://link.aps.org/doi/10.1103/PhysRevLett.94.123602}
\BIBentrySTDinterwordspacing

\bibitem{wallraff_approaching_2005}
\BIBentryALTinterwordspacing
A.~Wallraff \emph{et~al.}, ``Approaching unit visibility for control of a superconducting qubit with dispersive readout,'' \emph{Phys. Rev. Lett.}, vol.~95, no.~6, p. 060501, Aug. 2005. [Online]. Available: \url{https://link.aps.org/doi/10.1103/PhysRevLett.95.060501}
\BIBentrySTDinterwordspacing

\bibitem{huang_microwave_2021}
\BIBentryALTinterwordspacing
S.~Huang \emph{et~al.}, ``Microwave package design for superconducting quantum processors,'' \emph{PRX Quantum}, vol.~2, no.~2, p. 020306, Apr. 2021. [Online]. Available: \url{https://link.aps.org/doi/10.1103/PRXQuantum.2.020306}
\BIBentrySTDinterwordspacing

\bibitem{M.Tinkham_intro_sc}
M.~Tinkham, \emph{Introduction to superconductivity}, 2nd~ed.\hskip 1em plus 0.5em minus 0.4em\relax Mineola, NY, USA: Dover, 2004.

\bibitem{glazman_bogoliubov_2021}
\BIBentryALTinterwordspacing
L.~I. Glazman and G.~Catelani, ``Bogoliubov quasiparticles in superconducting qubits,'' \emph{SciPost Phys. Lect. Notes}, p.~31, Jun. 2021. [Online]. Available: \url{https://scipost.org/10.21468/SciPostPhysLectNotes.31}
\BIBentrySTDinterwordspacing

\bibitem{catelani_relaxation_2011}
\BIBentryALTinterwordspacing
G.~Catelani, R.~J. Schoelkopf, M.~H. Devoret, and L.~I. Glazman, ``Relaxation and frequency shifts induced by quasiparticles in superconducting qubits,'' \emph{Phys. Rev. B}, vol.~84, no.~6, p. 064517, Aug. 2011. [Online]. Available: \url{https://link.aps.org/doi/10.1103/PhysRevB.84.064517}
\BIBentrySTDinterwordspacing

\bibitem{aumentado_quasiparticle_2023}
\BIBentryALTinterwordspacing
J.~Aumentado, G.~Catelani, and K.~Serniak, ``Quasiparticle poisoning in superconducting quantum computers,'' \emph{Phys. Today}, vol.~76, no.~8, pp. 34--39, Aug. 2023. [Online]. Available: \url{https://doi.org/10.1063/PT.3.5291}
\BIBentrySTDinterwordspacing

\bibitem{wang_measurement_2014}
\BIBentryALTinterwordspacing
C.~Wang \emph{et~al.}, ``Measurement and control of quasiparticle dynamics in a superconducting qubit,'' \emph{Nat. Commun.}, vol.~5, no.~1, p. 5836, Dec. 2014. [Online]. Available: \url{https://www.nature.com/articles/ncomms6836}
\BIBentrySTDinterwordspacing

\bibitem{riste_millisecond_2013}
\BIBentryALTinterwordspacing
D.~Rist{\`e}, C.~C. Bultink, M.~J. Tiggelman, R.~N. Schouten, K.~W. Lehnert, and L.~DiCarlo, ``Millisecond charge-parity fluctuations and induced decoherence in a superconducting transmon qubit,'' \emph{Nat. Commun.}, vol.~4, no.~1, p. 1913, May 2013. [Online]. Available: \url{https://www.nature.com/articles/ncomms2936}
\BIBentrySTDinterwordspacing

\bibitem{somoroff_millisecond_2023}
\BIBentryALTinterwordspacing
A.~Somoroff, Q.~Ficheux, R.~A. Mencia, H.~Xiong, R.~Kuzmin, and V.~E. Manucharyan, ``Millisecond coherence in a superconducting qubit,'' \emph{Phys. Rev. Lett.}, vol. 130, no.~26, p. 267001, Jun. 2023. [Online]. Available: \url{https://link.aps.org/doi/10.1103/PhysRevLett.130.267001}
\BIBentrySTDinterwordspacing

\bibitem{bland_2d_2025}
\BIBentryALTinterwordspacing
M.~P. Bland \emph{et~al.}, ``2d transmons with lifetimes and coherence times exceeding 1 millisecond,'' 2025, arXiv:2503.14798. [Online]. Available: \url{https://arxiv.org/abs/2503.14798}
\BIBentrySTDinterwordspacing

\bibitem{tuokkola_methods_2025}
\BIBentryALTinterwordspacing
M.~Tuokkola \emph{et~al.}, ``Methods to achieve near-millisecond energy relaxation and dephasing times for a superconducting transmon qubit,'' Feb. 2025, arXiv:2407.18778. [Online]. Available: \url{https://arxiv.org/abs/2407.18778}
\BIBentrySTDinterwordspacing

\bibitem{gordon_environmental_2022}
\BIBentryALTinterwordspacing
R.~T. Gordon \emph{et~al.}, ``Environmental radiation impact on lifetimes and quasiparticle tunneling rates of fixed-frequency transmon qubits,'' \emph{Appl. Phys. Lett.}, vol. 120, no.~7, p. 074002, Feb. 2022. [Online]. Available: \url{https://doi.org/10.1063/5.0078785}
\BIBentrySTDinterwordspacing

\bibitem{connolly_coexistence_2024}
\BIBentryALTinterwordspacing
T.~Connolly \emph{et~al.}, ``Coexistence of nonequilibrium density and equilibrium energy distribution of quasiparticles in a superconducting qubit,'' \emph{Phys. Rev. Lett.}, vol. 132, no.~21, p. 217001, May 2024. [Online]. Available: \url{https://link.aps.org/doi/10.1103/PhysRevLett.132.217001}
\BIBentrySTDinterwordspacing

\bibitem{vepsalainen_impact_2020}
\BIBentryALTinterwordspacing
A.~P. Veps{\"a}l{\"a}inen \emph{et~al.}, ``Impact of ionizing radiation on superconducting qubit coherence,'' \emph{Nature}, vol. 584, no. 7822, pp. 551--556, Aug. 2020. [Online]. Available: \url{https://www.nature.com/articles/s41586-020-2619-8}
\BIBentrySTDinterwordspacing

\bibitem{mcewen_resolving_2022}
\BIBentryALTinterwordspacing
M.~McEwen \emph{et~al.}, ``Resolving catastrophic error bursts from cosmic rays in large arrays of superconducting qubits,'' \emph{Nat. Phys.}, vol.~18, no.~1, pp. 107--111, Jan. 2022. [Online]. Available: \url{https://www.nature.com/articles/s41567-021-01432-8}
\BIBentrySTDinterwordspacing

\bibitem{cardani_reducing_2021}
\BIBentryALTinterwordspacing
L.~Cardani \emph{et~al.}, ``Reducing the impact of radioactivity on quantum circuits in a deep-underground facility,'' \emph{Nat. Commun.}, vol.~12, no.~1, p. 2733, May 2021. [Online]. Available: \url{https://www.nature.com/articles/s41467-021-23032-z}
\BIBentrySTDinterwordspacing

\bibitem{harrington_synchronous_2024}
\BIBentryALTinterwordspacing
P.~M. Harrington \emph{et~al.}, ``Synchronous detection of cosmic rays and correlated errors in superconducting qubit arrays,'' Feb. 2024, arXiv:2402.03208. [Online]. Available: \url{https://arxiv.org/abs/2402.03208}
\BIBentrySTDinterwordspacing

\bibitem{wilen_correlated_2021}
\BIBentryALTinterwordspacing
C.~D. Wilen \emph{et~al.}, ``Correlated charge noise and relaxation errors in superconducting qubits,'' \emph{Nature}, vol. 594, no. 7863, pp. 369--373, Jun. 2021. [Online]. Available: \url{https://www.nature.com/articles/s41586-021-03557-5}
\BIBentrySTDinterwordspacing

\bibitem{yelton_correlated_2025}
\BIBentryALTinterwordspacing
E.~Yelton, C.~P. Larson, K.~Dodge, K.~Okubo, and B.~L.~T. Plourde, ``Correlated quasiparticle poisoning from phonon-only events in superconducting qubits,'' Mar. 2025, arXiv:2503.09554. [Online]. Available: \url{https://arxiv.org/abs/2503.09554}
\BIBentrySTDinterwordspacing

\bibitem{mohseni_how_2025}
\BIBentryALTinterwordspacing
M.~Mohseni \emph{et~al.}, ``How to build a quantum supercomputer: Scaling from hundreds to millions of qubits,'' Jan. 2025, arXiv:2411.10406. [Online]. Available: \url{https://arxiv.org/abs/2411.10406}
\BIBentrySTDinterwordspacing

\bibitem{mcewen_resisting_2024}
\BIBentryALTinterwordspacing
M.~McEwen \emph{et~al.}, ``Resisting high-energy impact events through gap engineering in superconducting qubit arrays,'' \emph{Phys. Rev. Lett.}, vol. 133, no.~24, p. 240601, Dec. 2024. [Online]. Available: \url{https://link.aps.org/doi/10.1103/PhysRevLett.133.240601}
\BIBentrySTDinterwordspacing

\bibitem{martinis_saving_2021}
\BIBentryALTinterwordspacing
J.~M. Martinis, ``Saving superconducting quantum processors from decay and correlated errors generated by gamma and cosmic rays,'' \emph{npj Quantum Information}, vol.~7, no.~1, pp. 1--9, Jun. 2021. [Online]. Available: \url{https://www.nature.com/articles/s41534-021-00431-0}
\BIBentrySTDinterwordspacing

\bibitem{aumentado_nonequilibrium_2004}
\BIBentryALTinterwordspacing
J.~Aumentado, M.~W. Keller, J.~M. Martinis, and M.~H. Devoret, ``Nonequilibrium quasiparticles and 2e periodicity in single-cooper-pair transistors,'' \emph{Phys. Rev. Lett.}, vol.~92, no.~6, p. 066802, Feb. 2004. [Online]. Available: \url{https://link.aps.org/doi/10.1103/PhysRevLett.92.066802}
\BIBentrySTDinterwordspacing

\bibitem{nho_recovery_2025}
\BIBentryALTinterwordspacing
H.~Nho \emph{et~al.}, ``Recovery dynamics of a gap-engineered transmon after a quasiparticle burst,'' May 2025, arXiv:2505.08104 [quant-ph]. [Online]. Available: \url{http://arxiv.org/abs/2505.08104}
\BIBentrySTDinterwordspacing

\bibitem{riwar_normal-metal_2016}
\BIBentryALTinterwordspacing
R.-P. Riwar \emph{et~al.}, ``Normal-metal quasiparticle traps for superconducting qubits,'' \emph{Phys. Rev. B}, vol.~94, no.~10, p. 104516, Sep. 2016. [Online]. Available: \url{https://link.aps.org/doi/10.1103/PhysRevB.94.104516}
\BIBentrySTDinterwordspacing

\bibitem{patel_phonon-mediated_2017}
\BIBentryALTinterwordspacing
U.~Patel, I.~V. Pechenezhskiy, B.~L.~T. Plourde, M.~G. Vavilov, and R.~McDermott, ``Phonon-mediated quasiparticle poisoning of superconducting microwave resonators,'' \emph{Phys. Rev. B}, vol.~96, no.~22, p. 220501, Dec. 2017. [Online]. Available: \url{https://link.aps.org/doi/10.1103/PhysRevB.96.220501}
\BIBentrySTDinterwordspacing

\bibitem{barends_minimizing_2011}
\BIBentryALTinterwordspacing
R.~Barends \emph{et~al.}, ``Minimizing quasiparticle generation from stray infrared light in superconducting quantum circuits,'' \emph{Appl. Phys. Lett.}, vol.~99, no.~11, p. 113507, Sep. 2011. [Online]. Available: \url{https://doi.org/10.1063/1.3638063}
\BIBentrySTDinterwordspacing

\bibitem{halpern_far_1986}
\BIBentryALTinterwordspacing
M.~Halpern, H.~P. Gush, E.~Wishnow, and V.~D. Cosmo, ``Far infrared transmission of dielectrics at cryogenic and room temperatures: Glass, fluorogold, eccosorb, stycast, and various plastics,'' \emph{Appl. Opt.}, vol.~25, no.~4, pp. 565--570, Feb. 1986. [Online]. Available: \url{https://opg.optica.org/ao/abstract.cfm?uri=ao-25-4-565}
\BIBentrySTDinterwordspacing

\bibitem{santavicca_impedance-matched_2008}
\BIBentryALTinterwordspacing
D.~F. Santavicca and D.~E. Prober, ``Impedance-matched low-pass stripline filters,'' \emph{Meas. Sci. Technol.}, vol.~19, no.~8, p. 087001, Jun. 2008. [Online]. Available: \url{https://dx.doi.org/10.1088/0957-0233/19/8/087001}
\BIBentrySTDinterwordspacing

\bibitem{paquette_absorptive_2022}
\BIBentryALTinterwordspacing
A.~Paquette \emph{et~al.}, ``Absorptive filters for quantum circuits: Efficient fabrication and cryogenic power handling,'' \emph{Appl. Phys. Lett.}, vol. 121, no.~12, p. 124001, Sep. 2022. [Online]. Available: \url{https://doi.org/10.1063/5.0114887}
\BIBentrySTDinterwordspacing

\bibitem{martinis_experimental_1987}
\BIBentryALTinterwordspacing
J.~M. Martinis, M.~H. Devoret, and J.~Clarke, ``Experimental tests for the quantum behavior of a macroscopic degree of freedom: The phase difference across a josephson junction,'' \emph{Phys. Rev. B}, vol.~35, no.~10, pp. 4682--4698, Apr. 1987. [Online]. Available: \url{https://link.aps.org/doi/10.1103/PhysRevB.35.4682}
\BIBentrySTDinterwordspacing

\bibitem{slichter_millikelvin_2009}
\BIBentryALTinterwordspacing
D.~H. Slichter, O.~Naaman, and I.~Siddiqi, ``Millikelvin thermal and electrical performance of lossy transmission line filters,'' \emph{Appl. Phys. Lett.}, vol.~94, no.~19, p. 192508, May 2009. [Online]. Available: \url{https://doi.org/10.1063/1.3133362}
\BIBentrySTDinterwordspacing

\bibitem{roy_introduction_2016}
\BIBentryALTinterwordspacing
A.~Roy and M.~Devoret, ``Introduction to parametric amplification of quantum signals with josephson circuits,'' \emph{Comptes Rendus Phys.}, vol.~17, no.~7, pp. 740--755, Aug. 2016. [Online]. Available: \url{https://www.sciencedirect.com/science/article/pii/S1631070516300640}
\BIBentrySTDinterwordspacing

\bibitem{aumentado_superconducting_2020}
\BIBentryALTinterwordspacing
J.~Aumentado, ``Superconducting parametric amplifiers: The state of the art in josephson parametric amplifiers,'' \emph{IEEE Microw. Mag.}, vol.~21, no.~8, pp. 45--59, Aug. 2020. [Online]. Available: \url{https://ieeexplore.ieee.org/document/9134828}
\BIBentrySTDinterwordspacing

\bibitem{pospieszalski_extremely_2005}
\BIBentryALTinterwordspacing
M.~W. Pospieszalski, ``Extremely low-noise amplification with cryogenic fets and hfets: 1970-2004,'' \emph{IEEE Microw. Mag.}, vol.~6, no.~3, pp. 62--75, Sep. 2005. [Online]. Available: \url{https://ieeexplore.ieee.org/document/1511915}
\BIBentrySTDinterwordspacing

\bibitem{zeng_sub-mw_2024}
\BIBentryALTinterwordspacing
Y.~Zeng, J.~Stenarson, P.~Sobis, N.~Wadefalk, and J.~Grahn, ``Sub-mw cryogenic inp hemt lna for qubit readout,'' \emph{IEEE Trans. Microw. Theory Techn.}, vol.~72, no.~3, pp. 1606--1617, Mar. 2024. [Online]. Available: \url{https://ieeexplore.ieee.org/document/10252156}
\BIBentrySTDinterwordspacing

\bibitem{krinner_engineering_2019}
\BIBentryALTinterwordspacing
S.~Krinner \emph{et~al.}, ``Engineering cryogenic setups for 100-qubit scale superconducting circuit systems,'' \emph{EPJ Quantum Technol.}, vol.~6, no.~1, pp. 1--29, Dec. 2019. [Online]. Available: \url{https://epjquantumtechnology.springeropen.com/articles/10.1140/epjqt/s40507-019-0072-0}
\BIBentrySTDinterwordspacing

\bibitem{simbierowicz_microwave_2022}
\BIBentryALTinterwordspacing
S.~Simbierowicz, V.~Y. Monarkha, S.~Singh, N.~Messaoudi, P.~Krantz, and R.~E. Lake, ``Microwave calibration of qubit drive line components at millikelvin temperatures,'' \emph{Appl. Phys. Lett.}, vol. 120, no.~5, p. 054004, Feb. 2022. [Online]. Available: \url{https://doi.org/10.1063/5.0081861}
\BIBentrySTDinterwordspacing

\bibitem{rehammar_low-pass_2023}
\BIBentryALTinterwordspacing
R.~Rehammar and S.~Gasparinetti, ``Low-pass filter with ultrawide stopband for quantum computing applications,'' \emph{IEEE Trans. Microw. Theory Techn.}, vol.~71, no.~7, pp. 3075--3080, Jul. 2023. [Online]. Available: \url{https://ieeexplore.ieee.org/document/10032269/citations?tabFilter=papers#citations}
\BIBentrySTDinterwordspacing

\bibitem{kurilovich_high-frequency_2025}
\BIBentryALTinterwordspacing
P.~D. Kurilovich \emph{et~al.}, ``High-frequency readout free from transmon multi-excitation resonances,'' Jan. 2025, arXiv:2501.09161. [Online]. Available: \url{https://arxiv.org/abs/2501.09161}
\BIBentrySTDinterwordspacing

\bibitem{hazra_benchmarking_2025}
\BIBentryALTinterwordspacing
S.~Hazra \emph{et~al.}, ``Benchmarking the readout of a superconducting qubit for repeated measurements,'' \emph{Phys. Rev. Lett.}, vol. 134, no.~10, p. 100601, Mar. 2025. [Online]. Available: \url{https://link.aps.org/doi/10.1103/PhysRevLett.134.100601}
\BIBentrySTDinterwordspacing

\bibitem{gaydamachenko_rf-squid-based_2025}
\BIBentryALTinterwordspacing
V.~Gaydamachenko, C.~Kissling, and L.~Grünhaupt, ``rf-squid-based traveling-wave parametric amplifier with input saturation power of -84 dbm across more than one octave in bandwidth,'' \emph{Phys. Rev. Appl.}, vol.~23, no.~6, p. 064053, Jun. 2025. [Online]. Available: \url{https://link.aps.org/doi/10.1103/1qk4-fzkq}
\BIBentrySTDinterwordspacing

\bibitem{pozar2012}
D.~M. Pozar, \emph{Microwave engineering}, 4th~ed.\hskip 1em plus 0.5em minus 0.4em\relax Hoboken, NJ, USA: Wiley, 2012.

\bibitem{comsol}
{COMSOL AB}, ``\textit{COMSOL Multiphysics\textsuperscript{\textregistered} v. 6.1},'' [Online]. Available: https://www.comsol.com, 2025.

\bibitem{2020SciPy-NMeth}
\BIBentryALTinterwordspacing
P.~Virtanen \emph{et~al.}, ``Scipy 1.0: Fundamental algorithms for scientific computing in python,'' \emph{Nat. Methods}, vol.~17, pp. 261--272, 2020. [Online]. Available: \url{https://rdcu.be/b08Wh}
\BIBentrySTDinterwordspacing

\end{thebibliography}

\end{document}